\documentclass[preprint,12pt]{elsart}

\usepackage{amssymb}
\usepackage{amsmath}
\usepackage{eepic}
\usepackage{amscd,graphicx}
\def\Z{{\mathbb Z}}
\def\R{{\mathbb R}}

\def\P{{\mathbb P}}
\def\T{{\mathbb T}}

\newcommand{\DIS}{\displaystyle}

\def\C{{\mathbb C}}
\def\F{{\mathbb F}}
\def\H{{\mathbb H}}

\def\bx{\text{\mathversion{bold}{$x$}}}

\newtheorem{theorem}{Theorem}
\newtheorem{corollary}{Corollary}
\newtheorem{remark}{Remark}

\journal{WAVES 2011}

\begin{document}

\begin{frontmatter}



\title{\bf{The group law on the tropical Hesse pencil}}


\author{Atsushi \textsc{Nobe}}

\address{Department of Mathematics, Faculty of Education, Chiba University,\\
1-33 Yayoi-cho Inage-ku, Chiba 263-8522, Japan}

\begin{abstract}
We show that the addition of points on the tropical Hesse curve can be realized via the intersection with a tropical line.
Then the addition formula for the tropical Hesse curve is reduced from those for the level-three theta functions through the ultradiscretization procedure.
A tropical analogue of the Hessian group $G_{216}$, the group of linear automorphisms acting on the Hesse pencil, is also investigated; it is shown that the dihedral group $\mathcal{D}_3$ of degree three is the group of linear automorphisms acting on the tropical Hesse pencil.
\end{abstract}


\end{frontmatter}

% \linenumbers


%-------------------------------------%
%-------------- SECTION --------------%
%-------------------------------------%
\section{Introduction}
The Hesse pencil is a canonical elliptic pencil consisting of the Hesse cubic curves to which all nonsingular cubic curves are projectively equivalent \cite{Hesse1844-1,Hesse1844-2}.
The Hessian curve\footnote{The Hessian curve $He(E)$ of a nonsingular cubic curve $E$ is the plane cubic curve defined by the equation $He(F) = 0$, where $He(F)$ is the determinant of the matrix of the second partial derivatives of the defining polynomial $F$ of $E$.} of the Hesse cubic curve is the curve itself, and therefore the nine base points of the Hesse pencil are its inflection points \cite{AD06}.
Moreover, the pencil is parametrized with the level-three theta functions, and hence it has the level-three structure \cite{AD06}.  
In addition to these remarkable properties, the Hesse pencil also plays an important role from the view point of tropical geometry; the tropical analogue of the Hesse pencil (the tropical Hesse pencil) is a pencil of tropical cubic curves of genus one and has a group structure in analogy to the Hesse pencil \cite{KKNT09,N11}.
This property contrasts well with the fact that the Weierstra$\ss$ canonical elliptic curve loses its group structure in the tropical limit.
The group structure of the tropical Hesse pencil confirms the existence of tropical analogues of the level-three theta functions parametrizing it.

It is well known that once we find a group structure of a plane curve then we can construct two kinds of dynamical systems realized as the evolutions of points on the curve; one is integrable and the other is solvable but chaotic \cite{QRT, Tsuda, KNT08, KKNT09, N11-2}.
Actually, we can construct such dynamical systems on the Hesse pencil as follows.
Let $P_0$ be an arbitrary point on the plane $\C^2$.
Then there exists a unique curve in the Hesse pencil passing through $P$.
We denote the curve by $E_\lambda$.
Fix a point $T$ on $E_{\lambda}$.
Let us consider the addition map 
\begin{align}
P_0\mapsto P_{1}=P_0+T.
\label{eq:QRT}
\end{align}
Then $P_1$ is a point on $E_\lambda$ because both $P_0$ and $T$ are on $E_\lambda$.
By applying the map repeatedly, we obtain the sequence of points $P_0,P_1,\ldots$ on $E_\lambda$.
This dynamical system is nothing but a member of the celebrated QRT system, a family of paradigmatic two-dimensional integrable maps \cite{QRT,Tsuda, N08,N11-2}. 
The parameter $\lambda$ of the pencil is the conserved quantity of the dynamical system, and the general solution to the dynamical system is concretely constructed  by using the level-three theta functions \cite{N08}.

On the other hand, if we consider the duplication map 
\begin{align*}
P_0\mapsto P_{1}=P_0+P_0=2P_0
%\label{eq:Dupl}
\end{align*}
instead of \eqref{eq:QRT} then we obtain the sequence of points $P_0,P_1,\ldots$ on $E_\lambda$ so called the solvable chaotic system \cite{Schroder1871,Umeno97,KNT08,KKNT09}.
This dynamical system behaves chaotically, nevertheless we can construct its general solution by using the level-three theta functions as in the case of the QRT system \cite{KNT08,KKNT09}.

In this paper, we review the group structure of the tropical Hesse pencil on which the above-mentioned types of piecewise linear map dynamical systems are constructed.
This group structure is geometrically realized as the intersection of the tropical Hesse curve with a tropical line.
It is also realized analytically as the ultradiscrete limit of the addition formulae for the level-three theta functions.  
In addition, we show that the group of linear automorphisms acting on the Hesse pencil (the Hessian group) survives as the dihedral group of degree three acting on the tropical Hesse pencil in the tropical limit \cite{N11-1}.

%-------------------------------------%
%-------------- SECTION --------------%
%-------------------------------------%
\section{Tropical Hesse pencil}
%---------------------%
% SUBSECTION %
%---------------------%
\subsection{Hesse pencil}
The Hesse pencil is a one-dimensional linear system of plane cubic curves in $\P^2(\C)$ given by the equation
\begin{align*}
f(x_0,x_1,x_2; t_0,t_1)
:=
t_0\left(x_0^3+x_1^3+x_2^3\right)+t_1 x_0x_1x_2
=
0,
\end{align*}
where $(x_0,x_1,x_2)$ is the homogeneous coordinate of $\P^2(\C)$ and the parameter $(t_0,t_1)$ ranges over $\P^1(\C)$ \cite{AD06}. 
Each curve composing the pencil is called the Hesse cubic curve and is denoted by $E_{t_0,t_1}$. %(see figure \ref{fig:HesseCurve}).
It is well known that every nonsingular cubic curve can be transformed projectively into a member of the Hesse pencil $\left\{E_{t_0,t_1}\right\}_{(t_0,t_1)\in\P^1(\C)}$ \cite{Hesse1844-1,Hesse1844-2}.
%Therefore the Hesse cubic curve is considered to be a canonical elliptic curve.
%//////////////////// FIGURE ////////////////////%
%\begin{figure}[htbp]
%\centering
%{
%\includegraphics[scale=.7]{HessePencil.eps}
%}
%\caption{Several members of the Hesse pencil. 
%}
%\label{fig:HesseCurve}
%\end{figure}
%//////////////////// FIGURE ////////////////////%

The nine  base points $p_0,p_1\ldots, p_8$ of the pencil are given as follows
\begin{align*}
&p_0=(0,1,-1),&
&p_1=(0,1,-\zeta_3),&
&p_2=(0,1,-\zeta_3^2),\\
&p_3=(1,0,-1),&
&p_4=(1,0,-\zeta_3^2),&
&p_5=(1,0,-\zeta_3),\\
&p_6=(1,-1,0),&
&p_7=(1,-\zeta_3,0),&
&p_8=(1,-\zeta_3^2,0),
\end{align*}
where $\zeta_3$ denotes the primitive third root of 1.
Any smooth curve in the pencil has these nine base points as its inflection points, and hence they are in the Hesse configuration 
\cite{AD06,SS31}.
%We choose $p_0$ as the unit of addition of the points on the Hesse cubic curve.

%---------------------%
% SUBSECTION %
%---------------------%
\subsection{Tropicalization of the Hesse pencil}
Let us tropicalize the Hesse pencil.
For the defining polynomial of the Hesse cubic curve we apply the procedure of tropicalization \cite{RST03,Gathmann06,IMS07}.
Replacing the operations $+$ and $\times$ with $\max$ and $+$ respectively, the defining polynomial $f(x_0,x_1,x_2; t_0,t_1)$ reduces to the tropical one
\begin{align*}
\tilde f(\tilde x_0,\tilde x_1,\tilde x_2; \tilde t_0,\tilde t_1)
:=
\max\left(
\tilde t_0+3\tilde x_0,\tilde t_0+3\tilde x_1,\tilde t_0+3\tilde x_2,\tilde t_1+\tilde x_0+\tilde x_1+\tilde x_2
\right).
\end{align*}
To distinguish the tropical variables form the original ones, they are ornamented with $\tilde{}\ $.

\begin{figure}[htbp]
\centering
{\unitlength=.035in{\def\arraystretch{1.0}
\begin{picture}(50,62)(-25,-30)
\thicklines
%%%%%%%%%% ab-axis %%%%%%%%%%
\put(0,28){\vector(0,1){2}}
\dottedline(0,-30)(0,28)
\put(0,32){\makebox(0,0){$Y$}}
\put(28,0){\vector(1,0){2}}
\dottedline(-30,0)(28,0)
\put(32,0){\makebox(0,0){$X$}}
\thicklines
%%%%%%%%%% curve %%%%%%%%%%
\put(-20,0){\line(1,-1){20}}
\put(0,-20){\line(1,2){20}}
\put(-20,0){\line(2,1){40}}
%%%%%%%%%% tentacle %%%%%%%%%%
\put(-20,0){\line(-1,0){10}}
\put(0,-20){\line(0,-1){10}}
\put(20,20){\line(1,1){10}}
%%%%%%%%%% label %%%%%%%%%%
\put(-3,-3){\makebox(0,0){$O$}}
\put(18,23){\makebox(0,0){$V_1$}}
\put(-20,4){\makebox(0,0){$V_2$}}
\put(4,-20){\makebox(0,0){$V_3$}}
\end{picture}
}}
\caption{
The tropical Hesse curve $C_\kappa$.
}
\label{fig:tropHesse}
\end{figure}
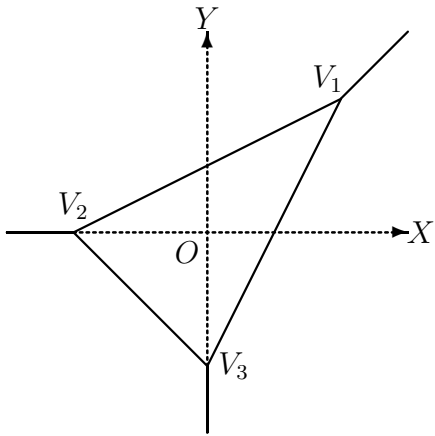

Let $(\tilde t_0,\tilde t_1)$ be a point in $\mathbb{TP}^1$, the tropical projective line.
Then $\tilde f$ can be regarded as a function $\tilde f:\mathbb{TP}^2\to\T$, where $\mathbb{TP}^2$ is the tropical projective plane and $\T:=\R\cup\{-\infty\}$ is the tropical semi-field \cite{MZ06}.
The tropicalization of the Hesse cubic curve is defined to be the set of points such that the function $\tilde f$ is not differentiable with respect to $\tilde x_1$, $\tilde x_2$, or $\tilde x_3$. 
We call the curve thus obtained the {tropical Hesse curve} and denote it by $C_{\tilde{t}_0,\tilde{t}_1}$.
Upon introduction of the inhomogeneous coordinate $(X,Y):=(\tilde{x}_1-\tilde{x}_0,\tilde{x}_2-\tilde{x}_0)\in\mathbb{TP}^2$ and $\kappa:=\tilde{t}_1-\tilde{t}_0\in\mathbb{TP}^1$ the tropical Hesse curve is given by the inhomogeneous tropical polynomial of degree three
\begin{align*}
F(X,Y; \kappa)
:=
\max\left(
3X,3Y,0,\kappa+X+Y
\right)
\end{align*}
and is denoted by $C_\kappa$ (see figure \ref{fig:tropHesse}).
%//////////////////// FIGURE ////////////////////%
%//////////////////// FIGURE ////////////////////%

The one-dimensional linear system $\{C_\kappa\}_{\kappa\in\mathbb{TP}^1}$ consisting of the tropical Hesse curves is called the tropical Hesse pencil.
The complement of the unbounded edges of $C_\kappa$ is denoted by $\bar C_\kappa:=C_\kappa\setminus\{\mbox{unbound edges}\}$. 
We denote the vertices whose coordinates are $(\kappa,\kappa)$, $(-\kappa,0)$, and $(0,-\kappa)$ by $V_1$, $V_2$, and $V_3$, respectively.
The three base points of the tropical Hesse pencil $\{C_\kappa\}_{\kappa\in\mathbb{TP}^1}$ are the end points of the unbounded edges emanating from $V_1$, $V_2$, and $V_3$, respectively.

We define the tropical Jacobian $J(C_\kappa)$ of the tropical Hesse curve $C_\kappa$ to be \cite{Vigeland04,MZ06,KKNT09}
\begin{align*}
J(C_\kappa)
:=
\R/3\kappa\Z
=
\{u\in \R\ |\ 0\leq u<3\kappa\}.
\end{align*}

%-------------------------------------%
%-------------- SECTION --------------%
%-------------------------------------%
\section{Addition of points on the tropical Hesse curve}
%---------------------%
% SUBSECTION %
%---------------------%
\subsection{Addition of points on the Hesse cubic curve}
Fix the parameter $(t_0,t_1)\in\P^1(\C)$ so that the Hesse cubic curve $E_{t_0,t_1}$ is nonsingular, or equivalently
\begin{align*}
(t_0,t_1)
\neq
(0,1),\
(1,-3),\
(1,-3\zeta_3),\
(1,-3\zeta_3^2).
\end{align*}
Let $K(E_{t_0,t_1})$ be the set of rational functions on the Hesse cubic curve $E_{t_0,t_1}$.
Let us consider the linear system with respect to the divisor $3p_0$ of degree three on $E_{t_0,t_1}$
\begin{align*}
L(3p_0)
:=
\left\{
h\in K(E_{t_0,t_1})\ |\ 
(h)+3p_0>0
\right\},
\end{align*}
where $(h)$ stands for the principal divisor of the rational function $h$ and $D>0$ means that $D$ is an effective divisor.

Consider the rational function $X_i$ ($i=0,1,2$) on $E_{t_0,t_1}$ such that $X_i(Q)=q_i$ for $Q=(q_0,q_1,q_2)\in E_{t_0,t_1}$.
Since $E_{t_0,t_1}$ is projectively equivalent to the Weierstra$\ss$  canonical form, the linear system $L(3p_0)$ is the three dimensional linear space spanned by the rational functions $X_0$, $X_1$, and $X_2$:
\begin{align*}
L(3p_0)
=
\left\langle
X_0,X_1,X_2
\right\rangle.
\end{align*}

Let $\mathcal{D}(E_{t_0,t_1})$ be the divisor group on $E_{t_0,t_1}$.
Let $\mathcal{D}_0(E_{t_0,t_1})$ be the subgroup of $\mathcal{D}(E_{t_0,t_1})$ generated by the divisors of degree 0.
Also let $\mathcal{D}_l(E_{t_0,t_1})$ be the group generated by the principal divisors on $E_{t_0,t_1}$.
Since $\deg (h)=0$ for any $h\in K(E_{t_0,t_1})$, we have $\mathcal{D}_l(E_{t_0,t_1})\subset \mathcal{D}_0(E_{t_0,t_1})$. 
There exists a bijection $\bar\phi$ from $E_{t_0,t_1}$ to the Jacobian $J(E_{t_0,t_1})\simeq{\rm Pic}^0(E_{t_0,t_1}):=\mathcal{D}_0(E_{t_0,t_1})/\mathcal{D}_l(E_{t_0,t_1})$ of $E_{t_0,t_1}$.
We can choose the bijection $\bar\phi: E_{t_0,t_1}\to J(E_{t_0,t_1})$ to be
\begin{align*}
\bar\phi(P)
\equiv
P-p_0
\quad
\mbox{(mod $\mathcal{D}_l(E_{t_0,t_1})$)}
\end{align*}
for $P\in E_{t_0,t_1}$.
The additive group structure of $E_{t_0,t_1}$ is induced from $J(E_{t_0,t_1})$ by $\bar\phi$.
Actually, the addition $P+Q$ of points $P,Q\in E_{t_0,t_1}$ is defined to be
\begin{align*}
P+Q
:=
\bar\phi(P)+\bar\phi(Q).
\end{align*}
Since $\bar\phi(p_0)=0$, the inflection point $p_0$ is the unit of addition.

Let $P$ and $Q$ be points on $E_{t_0,t_1}$.
If $P$ and $Q$ are in generic position then there exists a unique line $l_1$ passing through both $P$ and $Q$.
The line $l_1$ is the locus of all points $P$ such that $h(P)=0$ for $h\in L(3p_0)$.
Let the third intersection point of $E_{t_0,t_1}$ and $l_1$ be $\bar M$.
Then $P$, $Q$, and $\bar M$ satisfy $h(P)=h(Q)=h(\bar M)=0$, and hence $(h)=P+Q+\bar M-3p_0$ holds.
Thus we obtain
\begin{align*}
P+Q+\bar M
&=
\bar\phi(P)+\bar\phi(Q)+\bar\phi(\bar M)\\
&\equiv
P+Q+\bar M-3p_0\quad
\mbox{(mod $\mathcal{D}_l(E_{t_0,t_1})$)}\\
&\equiv
0\quad
\mbox{(mod $\mathcal{D}_l(E_{t_0,t_1})$)}\\
&=
p_0.
\end{align*}
Successively let $l_2$ be the line passing through both $p_0$ and $\bar M$.
Let the third intersection point of $E_{t_0,t_1}$ and $l_2$ be $M$.
Then we have
\begin{align*}
\bar M+M=p_0,
\end{align*}
and hence we obtain 
\begin{align}
P+Q=M.
\label{eq:addfm}
\end{align}
Thus the addition of points on $E_{t_0,t_1}$ is realized via the intersection with lines $l_1$ and $l_2$. %(see figure \ref{fig:HesseAdd}).

%//////////////////// FIGURE ////////////////////%
%\begin{figure}[htbp]
%\centering
%\includegraphics[scale=.3]{HesseAdd.pdf}
%\caption{Addition of points on the Hesse cubic curve.}
%\label{fig:HesseAdd}
%\end{figure}
%//////////////////// FIGURE ////////////////////%

%---------------------%
% SUBSECTION %
%---------------------%
\subsection{Theta functions}
Next we show that the addition of points on the Hesse cubic curve can also be realized via the addition formulae for the level-three theta functions.

The level-three theta functions $\theta_0(z,\tau)$, $\theta_1(z,\tau)$, and $\theta_2(z,\tau)$ are defined to be the theta functions with characteristics:
\begin{align*}
\theta_{k}(z,\tau)
:=
\vartheta_{\left(\frac{k}{3}-\frac{1}{6},\frac{3}{2}\right)}(3z,3\tau)
=
\sum_{n\in\Z}e^{3\pi i\left(n+\frac{k}{3}-\frac{1}{6}\right)^{2}\tau}
e^{6\pi i\left(n+\frac{k}{3}-\frac{1}{6}\right)\left(z+\frac{1}{2}\right)}
\end{align*}
for $k=0,1,2$, where $z\in\C$ and $\tau\in\H:=\{\tau\in\C\ |\ {\rm Im}\mkern2mu \tau>0\}$.
Fixing $\tau\in\H$, we abbreviate $\theta_k(z,\tau)$ and $\theta_k(0,\tau)$ as $\theta_k(z)$ and $\theta_k$, respectively.
We can easily see that the following holds
\begin{align}
&\theta_0=-\theta_1,
\qquad\theta_2=0.
\label{eq:thetaprop4}
\end{align}

Let $L_\tau:=(-\tau)\Z+(3\tau+1)\Z$ be the lattice in $\C$.
Consider the map $\varphi:\C\to\P^2(\C)$,
\begin{align*}
\varphi:\ z\longmapsto (\theta_2(z),\theta_0(z),\theta_1(z)).
\end{align*}
This induces an isomorphism from the complex torus $\C/L_\tau$ to the Hesse cubic curve $E_{{\theta_2^\prime},6{\theta_0^\prime}}$ \cite{KKNT09}.
It also induces the additive group structure on $E_{{\theta_2^\prime},6{\theta_0^\prime}}$ from $\C/L_\tau$ through the addition formulae for the level-three theta functions \cite{KKNT09}.
Note that the relation \eqref{eq:thetaprop4} implies the unit of addition on $E_{{\theta_2^\prime},6{\theta_0^\prime}}$ to be $p_0$:
\begin{align*}
\varphi:\ 0\longmapsto (\theta_2,\theta_0,\theta_1)=(0,1,-1)=p_0.
\end{align*}
%//////////////////// THEOREM ////////////////////%
\begin{theorem}\label{thm:addform}(See \cite{KKNT09})
Let the unit of addition on the Hesse cubic curve $E_{{\theta_2^\prime},6{\theta_0^\prime}}$ be $p_0$.
Let $P=(x_0,x_1,x_2)$ and $Q=(x_0^\prime,x_1^\prime,x_2^\prime)$ be points on $E_{{\theta_2^\prime},6{\theta_0^\prime}}$.
Then the addition $P+Q$ of the points is given as follows
\begin{align*}
P+Q
&=
(
x_1x_2{x_2^\prime}^2-x_0^2x_0^\prime x_1^\prime,
x_0x_1{x_1^\prime}^2-x_2^2x_0^\prime x_2^\prime,
x_0x_2{x_0^\prime}^2-x_1^2x_1^\prime x_2^\prime
)
\\
&=
(
x_0x_1{x_0^\prime}^2-x_2^2x_1^\prime x_2^\prime,
x_0x_2{x_2^\prime}^2-x_1^2x_0^\prime x_1^\prime,
x_1x_2{x_1^\prime}^2-x_0^2x_0^\prime x_2^\prime
)
\\
&=
(
x_0x_2{x_1^\prime}^2-x_1^2x_0^\prime x_2^\prime,
x_1x_2{x_0^\prime}^2-x_0^2x_1^\prime x_2^\prime,
x_0x_1{x_2^\prime}^2-x_2^2x_0^\prime x_1^\prime
).
\end{align*}
\end{theorem}
%//////////////////// THEOREM ////////////////////%
These formulae give the coordinate-wise expression of \eqref{eq:addfm}.

%---------------------%
% SUBSECTION %
%---------------------%
\subsection{Tropicalization of addition}
Fix the parameter $\kappa\in\mathbb{TP}^1$ so that the tropical Hesse curve $C_\kappa$ is nonsingular, or equivalently
\begin{align*}
0<\kappa<\infty.
\end{align*}
Let $K(C_\kappa)$ be the set of rational functions on $C_\kappa$.
A {rational function} on a tropical curve $E$ is a continuous function $h:E\to\mathbb{TP}^1$ such that the restriction of $h$ to any edge of $E$ is a piecewise linear integral function \cite{GK06}. 

The {order} ${\rm ord}_P h$ of a rational function $h$ at a point $P$ on a tropical curve $E$ is defined to be the sum of the outgoing slopes of all segments of $E$ emanating from $P$.
If ${\rm ord}_P h>0$ then the point $P$ is called the {zero} of $h$ of order ${\rm ord}_Ph$.
If ${\rm ord}_P h<0$ then $P$ is called the {pole} of $h$ of order $\left|{\rm ord}_Ph\right|$.
We define the {principal divisor} of a rational function $h$ on $E$ to be
\begin{align*}
(h)
:=
\sum_{P\in E}\left({\rm ord}_P h\right)P.
\end{align*}

Let us consider the set $K(\bar C_\kappa)$ of rational functions on $\bar C_\kappa$ (not $C_\kappa$).
We define the linear system $R(3V_1)$ with respect to the divisor $3V_1$ of degree three on $\bar C_\kappa$ as in the case of a non-tropical curve:
\begin{align*}
R(3V_1)
:=
\left\{
h\in K(\bar C_\kappa)\ |\ 
(h)+3V_1>0
\right\}.
\end{align*}

Consider the rational function $\tilde X_i$ ($i=1,2$) on $\bar C_\kappa$ such that $\tilde X_i(Q)=q_i$ for $Q=(q_1,q_2)\in \bar C_\kappa$.
Then the principal divisors of $\tilde X_i$ are given as follows
\begin{align*}
\left(\tilde X_1\right)
=
3V_2-3V_1,
\qquad
\left(\tilde X_2\right)
=
3V_3-3V_1.
\end{align*}
Therefore we have
\begin{align*}
\left(\tilde X_1\right)+3V_1=3V_2>0,
\qquad
\left(\tilde X_2\right)+3V_1=3V_1>0,
\end{align*}
and hence $\tilde X_1,\tilde X_2\in R(3V_1)$.
Consider the tropical module $\mathcal{M}$ generated by the rational functions 0, $\tilde X_1$, and $\tilde X_2$ \cite{MZ06}
\begin{align*}
\mathcal{M}
:=
\left\{
h\in K(\bar C_\kappa)\ |\
h
=
\max\left(
a,b+\tilde X_1,c+\tilde X_2
\right),\
a,b,c\in\T
\right\}.
\end{align*}
We  see that the following holds
\begin{align*}
\mathcal{M}
=
R(3V_1).
\end{align*}

Let $\mathcal{D}(\bar C_\kappa)$ be the divisor group on $\bar C_\kappa$.
Let $\mathcal{D}_0(\bar C_\kappa)$ be the subgroup of $\mathcal{D}(\bar C_\kappa)$ generated by the divisors of degree 0.
Also let $\mathcal{D}_l(\bar C_\kappa)$ be the group generated by the principal divisors on $\bar C_\kappa$.
Since $\deg (h)=0$ for any $h\in K(\bar C_\kappa)$, we have $\mathcal{D}_l(\bar C_\kappa)\subset \mathcal{D}_0(\bar C_\kappa)$.
There exists a bijection $\tilde\phi$ from $\bar C_\kappa$ (not $C_\kappa$) to the Jacobian $J(C_\kappa)\simeq{\rm Pic}^0(\bar C_\kappa)=\mathcal{D}_0(\bar C_\kappa)/\mathcal{D}_l(\bar C_\kappa)$ of $C_\kappa$ \cite{Vigeland04}.
We can choose the bijection $\tilde\phi: \bar C_\kappa\to J(C_\kappa)$ to be
\begin{align*}
\tilde\phi(P)
\equiv
P-V_1
\quad
\mbox{(mod $\mathcal{D}_l(\bar C_\kappa)$)}
\end{align*}
for $P\in \bar C_\kappa$.
Thus the additive group structure of $\bar C_\kappa$ is induced from $J(C_\kappa)$ by $\tilde\phi$.
Actually, the addition $P+Q$ of points $P,Q\in \bar C_\kappa$ is defined to be
\begin{align*}
P+Q
:=
\tilde\phi(P)+\tilde\phi(Q).
\end{align*}
Since $\bar\phi(V_1)=0$, the vertex $V_1$ is the unit of addition.

%//////////////////// REMARK ////////////////////%
\begin{remark}
If a point $P$ on $\bar C_\kappa$ is the zero of a rational function $h\in R(3V_1)$ then $h$ is not smooth at $P$ by definition.
Therefore the curve $B$ defined by $h$ passes through the point $P$.
Conversely, if $h$ passes through a point $P$ on $\bar C_\kappa$ then $P$ must be the zero of $h$ because $h$ is not smooth at $P$ and convex on the plane.
Thus the zero $P$ of $h$ is the intersection point of $\bar C_\kappa$ and $B$.
\end{remark}
%//////////////////// REMARK ////////////////////%

Let $P$ and $Q$ be points on $\bar C_\kappa$.
If $P$ and $Q$ are in generic position then there exists a unique line $r_1$ passing through both $P$ and $Q$.
The line $r_1$ is the locus of all points $P$ such that $h\in \mathcal{M}=R(3V_1)$ is not smooth at $P$.
Let the third intersection point of $\bar C_\kappa$ and $r_1$ be $\bar M$.
Then $P$, $Q$, and $\bar M$ satisfy $h(P)=h(Q)=h(\bar M)=0$, and hence $(h)=P+Q+\bar M-3V_0$ holds.
Thus we obtain
\begin{align*}
P+Q+\bar M
&=
\tilde\phi(P)+\tilde\phi(Q)+\tilde\phi(\bar M)\\
&\equiv
P+Q+\bar M-3V_1\quad
\mbox{(mod $\mathcal{D}_l(\bar C_\kappa)$)}\\
&\equiv
0\quad
\mbox{(mod $\mathcal{D}_l(\bar C_\kappa)$)}\\
&=
V_1.
\end{align*}
Successively let $r_2$ be the line passing through both $V_1$ and $\bar M$.
Let the third intersection point of $\bar C_\kappa$ and $r_2$ be $M$.
Then we have
\begin{align*}
\bar M+M=V_1,
\end{align*}
and hence we obtain 
\begin{align*}
P+Q=M.
%\label{eq:addfmtrop}
\end{align*}
Thus the addition of points on $\bar C_\kappa$ is realized via the intersection with lines $r_1$ and $r_2$. %(see figure \ref{fig:TropHesseAdd}).

%//////////////////// FIGURE ////////////////////%
%\begin{figure}[htbp]
%\begin{flushleft}
%\centering
% \includegraphics[scale=.3]{TropHesseAdd.pdf}
%\caption{Addition of points on the tropical Hesse curve.}
%\label{fig:TropHesseAdd}
%\end{flushleft}
%\end{figure}
%//////////////////// FIGURE ////////////////////%

%---------------------%
% SUBSECTION %
%---------------------%
\subsection{Ultradiscrete theta functions}
Next we show that the addition of points on the tropical Hesse curve can be derived from the addition formulae for the ultradiscrete theta functions. 
Since we choose the inhomogeneous coordinate of $\mathbb{TP}^1$ to describe the addition on the tropical Hesse curve, we rather consider the elliptic functions $c(z):={\theta_{0}(z,\tau)}/{\theta_{2}(z,\tau)}$ and $s(z):={\theta_{1}(z,\tau)}/{\theta_{2}(z,\tau)}$ and their ultradiscretizations denoted by $\tilde c(u)$ and $\tilde s(u)$.

Let $\kappa$ and $\varepsilon$ be positive numbers.
Let us fix the parameter $\tau$ in $\theta_k(z;\tau)$:
\begin{align*}
\tau
=
-\frac{3\kappa}{9\kappa+2\pi i \varepsilon}.
\end{align*}
Then the complex torus $\C/L_{\tau}$ converges into $J(C_\kappa)$ in the limit $\varepsilon\to0$ with respect to the Hausdorff metric \cite{N11}.

In order to ultradiscretize the level-three theta functions, we further assume 
\begin{align*}
z
=
\frac{\left(1-i\xi_\varepsilon \right)u}
{9\kappa}
\qquad
\mbox{and}
\qquad
 u\in\R,
\end{align*}
where $\xi_\varepsilon={2\pi\varepsilon}/{9\kappa}$, and take the limit $\varepsilon\to0$. 
Then we have  \cite{KKNT09,N11}
\begin{align*}
\lim_{\varepsilon\to0}\varepsilon \log c(z)
=
\tilde c(u),
\qquad
\lim_{\varepsilon\to0}\varepsilon \log s(z)
=
\tilde s(u),
\end{align*}
where we define
\begin{align*}
&\tilde c(u)
:=
-\frac{9\kappa}{2}
\left\{
\left(\left(
\frac{u-\kappa}{3\kappa}-\frac{1}{2}
\right)\right)
\right\}^2
+
\frac{9\kappa}{2}
\left\{
\left(\left(
\frac{u-3\kappa}{3\kappa}-\frac{1}{2}
\right)\right)
\right\}^2\\
&\tilde s(u)
:=
-\frac{9\kappa}{2}
\left\{
\left(\left(
\frac{u-2\kappa}{3\kappa}-\frac{1}{2}
\right)\right)
\right\}^2
+
\frac{9\kappa}{2}
\left\{
\left(\left(
\frac{u-3\kappa}{3\kappa}-\frac{1}{2}
\right)\right)
\right\}^2,
\end{align*}
and $\left(\left(u\right)\right):=u-{\rm Floor}(u)$.
These piecewise linear functions $\tilde c(u)$ and $\tilde s(u)$ are periodic with period $3\kappa$.

Now take the following representatives $z_{0k},z_{k1},z_{k2}$ of the zeros of the level-three theta functions $\theta_k(z)$ in $\C/L_\tau$ for $k=0,1,2$
\begin{align*}
\left(
\begin{matrix}
z_{20}&z_{21}&z_{22}\\[5pt]
z_{00}&z_{01}&z_{02}\\[5pt]
z_{10}&z_{11}&z_{12}\\[5pt]
\end{matrix}
\right)
=
\left(
\begin{matrix}
0&\tau+\frac{1}{3}&2\tau+\frac{2}{3}\\[5pt]
-\frac{\tau}{3}&\frac{2\tau}{3}+\frac{1}{3}&\frac{5\tau}{3}+\frac{2}{3}\\[5pt]
-\frac{2\tau}{3}&\frac{\tau}{3}+\frac{1}{3}&\frac{4\tau}{3}+\frac{2}{3}\\[5pt]
\end{matrix}
\right).
%\label{eq:thetazero}
\end{align*}
These nine zeros of the level-three theta functions are mapped into the nine inflection points on $E_{{\theta_2^\prime},6{\theta_0^\prime}}$ by  $\varphi$, respectively:
\begin{align*}
\varphi:\quad
\begin{matrix}
z_{20}&z_{21}&z_{22}\\
z_{00}&z_{01}&z_{02}\\
z_{10}&z_{11}&z_{12}\\
\end{matrix}
\quad
\longmapsto
\quad
\begin{matrix}
p_0&p_1&p_2\\
p_3&p_4&p_5\\
p_6&p_7&p_8\\
\end{matrix}\ .
%\label{eq:ztop}
\end{align*}

In terms of the variable $u$, we put the limit of zeros $z_{kj}$ $(k,j=0,1,2)$ of the level-three theta functions as follows
\begin{align*}
&u_2:=\lim_{\varepsilon\to0}9\kappa z_{20}=\lim_{\varepsilon\to0}9\kappa z_{21}=\lim_{\varepsilon\to0}9\kappa z_{22}=0,
\\
&u_0:=\lim_{\varepsilon\to0}9\kappa z_{00}=\lim_{\varepsilon\to0}9\kappa z_{01}=\lim_{\varepsilon\to0}9\kappa z_{02}=\kappa ,
\\
&u_1:=\lim_{\varepsilon\to0}9\kappa z_{10}=\lim_{\varepsilon\to0}9\kappa z_{11}=\lim_{\varepsilon\to0}9\kappa z_{12}=2\kappa ,
\end{align*}
where it should be noted that $\tau\to-1/3$ in the limit $\varepsilon\to0$. 

Let us introduce the map $\tilde\varphi: J(C_\kappa)\to \R^2\subset\mathbb{TP}^2$
\begin{align*}
\tilde\varphi:\quad
u\longmapsto(\tilde c(u),\tilde s(u)).
\end{align*}
The map $\tilde\varphi$ induces an isomorphism $\bar C_\kappa\simeq J(C_\kappa)$ \cite{KKNT09,N11}.
Thus $\tilde\varphi$ induces the additive group structure on $\bar C_\kappa$ equipped with the unit of addition $V_1=\tilde\varphi(0)$  from $J(C_\kappa)$.

Consider the map $\eta:E_{{\theta_2^\prime},6{\theta_0^\prime}}\to \bar C_\kappa $ so defined that the diagram commutes
\begin{align*}
\begin{CD}
\C/L_\tau
@ > \varepsilon\to0 >> 
J(C_\kappa )\\
@ V \varphi VV
@ VV \tilde\varphi V\\
E_{{\theta_2^\prime},6{\theta_0^\prime}}
@ > \eta >>
\bar C_\kappa .\\
\end{CD}
\end{align*}
The inflection points of $E_{{\theta_2^\prime},6{\theta_0^\prime}}$ are mapped into the vertices of $\bar C_\kappa $ by $\eta$:
\begin{align}
&\eta:\ 
p_0,\ p_1,\ p_2
\overset{\varphi^{-1}}{\longmapsto}
z_{20},\ z_{21},\ z_{22}
\overset{\varepsilon\to0}{\longrightarrow}
u_2
\overset{\tilde\varphi}{\longmapsto}
V_1,
\label{eq:eta1}\\
&\eta:\ 
p_3,\ p_4,\ p_5
\overset{\varphi^{-1}}{\longmapsto}
z_{00},\ z_{01},\ z_{02}
\overset{\varepsilon\to0}{\longrightarrow}
u_0
\overset{\tilde\varphi}{\longmapsto}
V_2,
\label{eq:eta2}\\
&\eta:\ 
p_6,\ p_7,\ p_8
\overset{\varphi^{-1}}{\longmapsto}
z_{10},\ z_{11},\ z_{12}
\overset{\varepsilon\to0}{\longrightarrow}
u_1
\overset{\tilde\varphi}{\longmapsto}
V_3.
\label{eq:eta3}
\end{align}
Note that the unit $p_0$ of addition on the Hesse cubic curve is mapped into $V_0$, the unit of addition on the tropical Hesse curve, by $\eta$.

Define the open subsets ${D}_1$, ${D}_2$, and ${D}_3$ of $J(C_\kappa)$ to be
\begin{align*}
D_j
:=
\left\{
u\in J(C_\kappa)\ |\ {(j-1)\kappa}< u<{j\kappa}
\right\}
\qquad
(j=1,2,3).
\end{align*}
Then we have $J(C_\kappa)=\bigcup_{j=0}^2 \left(D_{j+1}\cup u_j\right)$.
The addition formula for $\bar C_\kappa$ is explicitly given as follows.
In this theorem we denote $\max(\ )$ simply by $(\ )$.

%//////////////////// THEOREM ////////////////////%
\begin{theorem}\label{thm:adformuellip}(See \cite{N11})
Assume $u\in\overline{D_j}$ for a fixed $j=1,2,3$, where $\overline{D_j}$ is the closure of ${D_j}$.
Then the ultradiscrete elliptic functions $\tilde c$ and $\tilde s$ satisfy the following addition formulae
\begin{subequations}
\begin{align*}
&\tilde c(u+v)
=
\left(
\tilde s(u),2\tilde c(u)+\tilde c(v)+\tilde s(v)
\right)
-
\left(
\tilde c(u)+2\tilde c(v),2\tilde s(u)+\tilde s(v)
\right),
%\label{eq:uaddelliptic1a}
\\
&\tilde s(u+v)
=
\left(
\tilde c(u)+\tilde s(u)+2\tilde s(v),\tilde c(v)
\right)
-
\left(
\tilde c(u)+2\tilde c(v),2\tilde s(u)+\tilde s(v)
\right),
%\label{eq:uaddelliptic1b}
\end{align*}
\end{subequations}
if and only if $v\in \overline{D_j\cup D_{j+1}}$, 
\begin{subequations}
\begin{align*}
&\tilde c(u+v)
=
\left(
\tilde c(u)+\tilde s(u)+2\tilde c(v),\tilde s(v)
\right)
-
\left(
\tilde s(u)+2\tilde s(v),2\tilde c(u)+\tilde c(v)
\right),
%\label{eq:uaddelliptic2a}
\\
&\tilde s(u+v)
=
\left(
\tilde c(u),2\tilde s(u)+\tilde c(v)+\tilde s(v)
\right)
-
\left(
\tilde s(u)+2\tilde s(v),2\tilde c(u)+\tilde c(v)
\right),
%\label{eq:uaddelliptic2b}
\end{align*}
\end{subequations}
if and only if $v\in \overline{D_j\cup D_{j+2}}$, or
\begin{subequations}
\begin{align*}
&\tilde c(u+v)
=
\left(
\tilde c(u)+2\tilde s(v),2\tilde s(u)+\tilde c(v)
\right)
-
\left(
\tilde c(u)+\tilde s(u),\tilde c(v)+\tilde s(v)
\right),
%\label{eq:uaddelliptic3a}
\\
&\tilde s(u+v)
=
\left(
\tilde s(u)+2\tilde c(v),2\tilde c(u)+\tilde s(v)
\right)
-
\left(
\tilde c(u)+\tilde s(u),\tilde c(v)+\tilde s(v)
\right),
%\label{eq:uaddelliptic3b}
\end{align*}
\end{subequations}
if and only if $v\in \overline{D_{j+1}\cup D_{j+2}}$, where the subscripts are reduced modulo 3.
\end{theorem}
%//////////////////// THEOREM ////////////////////%

It immediately follows the addition formula for the points on the tropical Hesse curve $C_\kappa $ from theorem \ref{thm:adformuellip}.
Let the edge of $C_\kappa$ connecting the vertex $V_i$ with $V_{i+1}$ be $E_i$ for $i=1,2,3$, where we assume $V_4=V_1$.
Then we have the following corollary of theorem \ref{thm:adformuellip}.
%//////////////////// COROLLARY ////////////////////%
\begin{corollary}
\label{cor:adformuellip}
Let $P=(X,Y)$ be a point on an edge $E_j$ of the tropical Hesse curve $C_\kappa$ for a fixed $j=1,2,3$.
Then the point $P+ Q=(X+ X^\prime, Y+ Y^\prime)$ is given by the following addition formulae
\begin{subequations}
\begin{align*}
&X+ X^\prime
=
\max\left(
Y,2X+X^\prime+Y^\prime
\right)
-
\max\left(
X+2X^\prime,2Y+Y^\prime
\right),
%\label{eq:uaddHesse1a}
\\
&Y+ Y^\prime
=
\max\left(
X+Y+2Y^\prime,X^\prime
\right)
-
\max\left(
X+2X^\prime,2Y+Y^\prime
\right),
%\label{eq:uaddHesse1b}
\end{align*}
\end{subequations}
if and only if $Q=(X^\prime,Y^\prime)\in E_j\cup E_{j+1}$, 
\begin{subequations}
\begin{align*}
&X+ X^\prime
=
\max\left(
X+Y+2X^\prime,Y^\prime
\right)
-
\max\left(
Y+2Y^\prime,2X+X^\prime
\right),
%\label{eq:uaddHesse2a}
\\
&Y+ Y^\prime
=
\max\left(
X,2Y+X^\prime+Y^\prime
\right)
-
\max\left(
Y+2Y^\prime,2X+X^\prime
\right),
%\label{eq:uaddHesse2b}
\end{align*}
\end{subequations}
if and only if $Q\in E_j\cup E_{j+2}$, or
\begin{subequations}
\begin{align*}
&X+ X^\prime
=
\max\left(
X+2Y^\prime,2Y+X^\prime
\right)
-
\max\left(
X+Y,X^\prime+Y^\prime
\right),
%\label{eq:uaddHesse3a}
\\
&Y+ Y^\prime
=
\max\left(
Y+2X^\prime,2X+Y^\prime
\right)
-
\max\left(
X+Y,X^\prime+Y^\prime
\right),
%\label{eq:uaddHesse3b}
\end{align*}
\end{subequations}
if and only if $Q\in E_{j+1}\cup E_{j+2}$, where the subscripts are reduced modulo 3.
\end{corollary}
%//////////////////// COROLLARY ////////////////////%

%-------------------------------------%
%-------------- SECTION --------------%
%-------------------------------------%
\section{A tropical analogue of the Hessian group}\label{sec:HG216}
%---------------------%
% SUBSECTION %
%---------------------%
\subsection{Hessian group}
The Hessian group $G_{216}\simeq \Gamma\rtimes SL(2,\F_3)$ is a subgroup of $PGL(3,\C)$, where $\Gamma=\left(\Z/3\Z\right)^2$ and $SL(2,\F_3)$ is the special linear group over the finite field $\F_3$ of characteristic three.
The Hessian group is generated by the following four linear transformations
\begin{align*}
g_1
=
\left(\begin{matrix}
0&1&0\\
0&0&1\\
1&0&0\\
\end{matrix}\right),
\qquad
g_2
=
\left(\begin{matrix}
1&0&0\\
0&\zeta_3&0\\
0&0&\zeta_3^2\\
\end{matrix}\right),
\qquad
g_3
=
\left(\begin{matrix}
1&1&1\\
1&\zeta_3&\zeta_3^2\\
1&\zeta_3^2&\zeta_3\\
\end{matrix}\right),
\qquad
g_4
=
\left(\begin{matrix}
1&0&0\\
0&\zeta_3&0\\
0&0&\zeta_3\\
\end{matrix}\right).
\end{align*}
The name, ``Hessian" group, comes from the fact that $G_{216}$ is the group of linear automorphisms acting on the Hesse pencil \cite{AD06,SS31}.

The group of three torsion points on $E_{t_0,t_1}$, denoted by $E_{t_0,t_1}[3]$, consists of its nine inflection points $p_0,p_1,\ldots,p_8$.
The map
\begin{align*}
p_1 \longmapsto (1,0)
\qquad
p_3 \longmapsto (0,1)
\end{align*}
induces a group isomorphism from $E_{t_0,t_1}[3]$ to the normal subgroup $\Gamma=\left\langle g_1,g_2\right\rangle$ of $G_{216}$.
Therefore, the action of $\Gamma$ fixes the parameter $(t_0,t_1)$ of the pencil.

Let $\alpha: G_{216}\to PGL(2,\C)$ be the map given by
\begin{align*}
\alpha(g):\ 
(t_0,t_1)
=
(x_0x_1x_2, x_0^3+x_1^3+x_2^3)
\longmapsto
(t_0^\prime,t_1^\prime)
=
(x_0^\prime x_1^\prime x_2^\prime , x_0^{\prime3}+x_1^{\prime3}+x_2^{\prime3}),
\end{align*}
where $g\in G_{216}$ and $g: (x_0, x_1, x_2)\mapsto(x_0^\prime, x_1^\prime, x_2^\prime)$.
Then we have ${\rm Ker}(\alpha)\supset\Gamma$,
while $\alpha(g_3)$ and $\alpha(g_4)$ act effectively on $\P^1(\C)$: 
\begin{align*}
&\alpha(g_3):\ 
(t_0,t_1)
\longmapsto
(t_0^\prime,t_1^\prime)
=
(3t_0+t_1,18t_0-3t_1)\\
&\alpha(g_4):\ 
(t_0,t_1)
\longmapsto
(t_0^\prime,t_1^\prime)
=
(t_0,\zeta_3^2t_1).
\end{align*}
Thus $g_3$ and $g_4$ induce the action on the Hesse pencil independent of its additive group structure on each curve.
We can easily see that $\alpha(g_3)^2=\alpha(g_4)^3=1$ holds.
It follows that we have
\begin{align*}
\alpha(G_{216})
=
\left\langle
\alpha(g_3),\alpha(g_4)
\right\rangle
\simeq
\mathcal{T},
\end{align*}
where $\mathcal{T}$ is the tetrahedral group.
Thus the group $\alpha(G_{216})$ acts on $\P^1(\C)$ as the permutations among the following 12 points
\begin{align*}
&\lambda:=\frac{t_1}{t_0},\
\zeta_3\lambda,\
\zeta_3^2\lambda,\
\frac{18-3\lambda}{3+\lambda},\
\frac{18\zeta_3-3\zeta_3\lambda}{3+\lambda},\
\frac{18\zeta_3^2-3\zeta_3^2\lambda}{3+\lambda},\
\frac{18-3\zeta_3\lambda}{3+\zeta_3\lambda},\\
&\frac{18\zeta_3-3\zeta_3^2\lambda}{3+\zeta_3\lambda},\
\frac{18\zeta_3^2-3\lambda}{3+\zeta_3\lambda},\
\frac{18-3\zeta_3^2\lambda}{3+\zeta_3^2\lambda},\
\frac{18\zeta_3-3\lambda}{3+\zeta_3^2\lambda},\
\frac{18\zeta_3^2-3\zeta_3\lambda}{3+\zeta_3^2\lambda}.
\end{align*}

The Hesse pencil contains four singular members each of which is with multiplicity three.
These singular members correspond to the following four points in $\P^1(\C)$, respectively \cite{AD06}
\begin{align*}
(t_0,t_1)
=
(0,1),\ 
(1,-3),\ 
(1,-3\zeta_3^2),\ 
(1,-3\zeta_3).
\end{align*}
Denote these points by $s_i$ ($i=1,2,3,4$) in order.
These $s_i$'s are permuted by $\alpha(G_{216})$ as follows
\begin{align*}
&\alpha(g_3):\ 
s_1\longleftrightarrow s_2,
\quad
s_3\longleftrightarrow s_4
%\label{eq:actalpg3}
\\
&\alpha(g_4):\ 
s_2\longrightarrow s_3\longrightarrow s_4\longrightarrow s_2
\quad
\mbox{($s_1$ is fixed.)}
%\label{eq:actalpg4}
\end{align*}
%Thus $s_i$'s can be corresponded to the vertices of the tetrahedron on which $\mathcal{T}\simeq \alpha(G_{216})$ acts. 

Let
\begin{align*}
g_0
=
\left(\begin{matrix}
1&0&0\\
0&0&1\\
0&1&0\\
\end{matrix}\right).
\end{align*}
Then we have
\begin{align*}
g_3^2
=
\left(g_4g_0\right)^3
=g_0.
\end{align*}
Therefore, we obtain
\begin{align*}
G_{216}/\Gamma
=
\left\langle
g_3,g_4
\right\rangle
\simeq
\tilde{\mathcal{T}},
\end{align*}
where $\tilde{\mathcal{T}}$ is the binary tetrahedral group.
Since $\tilde{\mathcal{T}}$ is isomorphic to $SL(2,\F_3)$, we obtain the semi-direct product decomposition $G_{216}\simeq \Gamma\rtimes SL(2,\F_3)$.

Note that the actions of $g_1$ and $g_2$ on $E_{{\theta_2^\prime},6{\theta_0^\prime}}$ can be realized as the additions with $p_6$ and $p_1$, respectively
\begin{align*}
&(x_0,x_1,x_2)
\quad
\overset{g_1}{\longmapsto}
\quad
(x_1,x_2,x_0)
=
(x_0,x_1,x_2)+p_6,\\
&(x_0,x_1,x_2)
\quad
\overset{g_2}{\longmapsto}
\quad
(x_0,\zeta_3 x_1,\zeta_3^2 x_2)
=
(x_0,x_1,x_2)+p_1.
\end{align*}

%---------------------%
% SUBSECTION %
%---------------------%
\subsection{Tropicalization of the Hessian group}
Now we investigate the tropical analogue of the Hessian group $G_{216}\simeq \Gamma\rtimes SL(2,\F_3)$.
At first we consider the normal subgroup $\Gamma\simeq\left(\Z/3\Z\right)^2$.
Note that $\Gamma=\langle g_1,g_2\rangle$ and the actions of $g_1$ and $g_2$ on $E_{t_0,t_1}$ is realized as the additions with $p_6$ and $p_1$, respectively.

The correspondence \eqref{eq:eta1}, \eqref{eq:eta2}, and \eqref{eq:eta3} in terms of $\eta$ tell us that the addition with $p_6$ corresponds to that with $V_3$ on $\bar C_\kappa$, while that with $p_1$ vanishes in the limit $\varepsilon\to0$.
 (Note that  $V_1$ is the unit of addition on $\bar C_\kappa$.)
Since the addition with $p_3$ (resp. $V_2$) is equivalent to that with $2p_6$ (resp. $2V_3$), the tropical analogue of $\Gamma$ consists of the addition with $V_3$. 
Actually, it is the group $\bar C_\kappa[3]=\left\langle V_3\right\rangle\simeq Z/3\Z$ of three torsion points on $\bar C_\kappa$.
Denote the tropical analogue of a group $G$ by $trop(G)$.
Then we have
\begin{align*}
trop(\Gamma)
\simeq
\Z/3\Z.
\end{align*}
The addition of points $(X,Y)$ on $\bar C_\kappa$ with $V_3$ is explicitly computed as follows
\begin{align*}
(X,Y)+ V_3
=
(X,Y)+ (0,-\kappa)
=
(Y-X,-X),
\end{align*}
where we apply the addition formula in corollary \ref{cor:adformuellip}.

The group $trop(\Gamma)$ can also be obtained by applying the procedure of ultradiscretization directly to $g_1$ and $g_2$.
Let us consider the inhomogeneous coordinate $(x:=x_1/x_0,y:=x_2/x_0)$ of $\P^2(\C)$.
Let $g_1:(x,y)\mapsto (x^\prime,y^\prime)$ and $g_2:(x,y)\mapsto (x^{\prime\prime},y^{\prime\prime})$.
Then we have
\begin{align*}
\left(\left|x^\prime\right|,\left|y^\prime\right|\right)
=
\left(
\frac{|y|}{|x|},\frac{1}{|x|}
\right),
\qquad
\left(\left|x^{\prime\prime}\right|,\left|y^{\prime\prime}\right|\right)
=
\left(
|x|,|y|
\right).
\end{align*}
Replace $|x|$ and $|y|$ with $e^{X/\varepsilon}$ and $e^{Y/\varepsilon}$ for $X,Y\in\R$ and $\varepsilon>0$, respectively.
If we take the limit $\varepsilon\to0$ then we obtain
\begin{align*}
(X,Y)
\overset{\tilde g_1}{\longmapsto}
(Y-X,-X)
=
(X,Y)+ V_3,
\qquad
(X,Y)
\overset{\tilde g_2}{\longmapsto}
(X,Y),
\end{align*} 
where we denote the action on $\mathbb{TP}^2$ induced form $g_1$ and $g_2$ by $\tilde g_1$ and $\tilde g_2$, respectively.

Next we consider $\alpha(G_{216})\simeq{\mathcal{T}}$.
Consider the singular members of the tropical Hesse pencil $C_{\infty}$ and $C_0$, each of which is the tropicalization of the singular members $E_{s_1}$ or $E_{s_i}$, ($i=2,3,4$) of the Hesse pencil \cite{N11}.
Then the action of $\alpha(g_3)$, which permutes $s_1$ and $s_3$ with $s_2$ and $s_4$ respectively, must vanish; while the action of $\alpha(g_4)$, which fixes $s_1$ and permutes $s_2$, $s_3$, and $s_4$ cyclically, reduces to the action fixing both $C_0$ and $C_\infty$.
Therefore, we conclude
\begin{align*}
trop\left(\alpha(G_{216})\right)
\simeq
trop\left(\mathcal{T}\right)
\simeq
\left\langle 1\right\rangle.
\end{align*}
Thus the tropical analogue of the Hessian group fixes each member of the tropical Hesse pencil.

Furthermore, we consider the tropicalization of the element $g_0=g_3^2$ of $G_{216}/\Gamma$.
We ultradiscretize $g_0$ directly as well as $g_1$ and $g_2$.
In the inhomogeneous coordinate, the action of $g_0$ on $\P^2(\C)$ is simply
\begin{align*}
\left(
x,y
\right)
\overset{g_0}{\longmapsto}
\left(
{y},{x}
\right).
\end{align*}
It follows that we have
\begin{align*}
\left(
X,Y
\right)
\overset{\tilde g_0}{\longmapsto}
\left(
{Y},{X}
\right),
\end{align*}
where $(X,Y)$ is the inhomogeneous coordinate of $\mathbb{TP}^2$ and $\tilde g_0$ is the action on $\mathbb{TP}^2$ induced from $g_0$ by applying the procedure of ultradiscretization.
Thus we conclude that the tropical analogue of $\tilde{\mathcal{T}}\simeq\langle g_3,g_4\rangle=G_{216}/\Gamma$ is the group of order two generated by $\tilde g_0$:
\begin{align*}
trop\left(\tilde{\mathcal{T}}\right)
\simeq
\left\langle \tilde g_0\right\rangle
\simeq
\left\langle \left(\begin{matrix}2&0\\0&2\\\end{matrix}\right)\right\rangle
\subset SL(2,\F_3).
\end{align*}
%where $\left(\begin{matrix}2&0\\0&2\\\end{matrix}\right)\in SL(2,\F_3)$.

We finally obtain the following theorem concerning the tropical analogue of the Hessian group $G_{216}$.
%//////////////////// THEOREM ////////////////////%
\begin{theorem}
The dihedral group $\mathcal{D}_3$ of degree three,
\begin{align*}
\mathcal{D}_3
=\left\langle \tilde g_0, \tilde g_1\right\rangle
\simeq
\left(\Z/3\Z\right)
\rtimes
\left\langle
\left(
\begin{matrix}
2&0\\
0&2\\
\end{matrix}
\right)
\right\rangle
\end{align*}
where $\tilde g_0, \tilde g_1\in PGL(3,\T)$ satisfy $\tilde g_0^2 =\tilde g_1^3=\left(\tilde g_0\tilde g_1\right)^2=1$, is the group of linear automorphisms acting on the tropical Hesse pencil.
The action of $\tilde g_0$ on each curve of the pencil is realized as the reflection with respect to the line $Y=X$ passing through the vertex $V_1$; and the action of $\tilde g_1$ on each curve is realized as the addition with $V_3$.
\end{theorem}
%//////////////////// THEOREM ////////////////////%

The semi-direct product in $\mathcal{D}_3$ is defined as follows.
Let 
\begin{align*}
\gamma_1,\gamma_2\in\Z/3\Z
\qquad\mbox{and}\qquad
m_1,m_2\in\langle\left(\begin{matrix}2&0\\0&2\\ \end{matrix}\right)\rangle.
\end{align*}
Define the multiplication of  $(\gamma_1,m_1), (\gamma_2,m_2)\in\mathcal{D}_3$ to be
\begin{align*}
(\gamma_1,m_1)\cdot(\gamma_2,m_2)
=
(\gamma_1+m_1\gamma_2, m_1m_2),
\end{align*}
where we identify $\gamma_i$ with $(\gamma_i,0)\in\left(\Z/3\Z\right)^2$.

%-------------------------------------%
%-------------- SECTION --------------%
%-------------------------------------%
\section{Conclusion}
We gave two realizations of the addition of points on the tropical Hesse curve.
The one realized as the intersection of the tropical Hesse curve with a tropical line is reduced from the divisor class group and the linear system on the curve; the other realized as the addition formula for the ultradiscrete theta functions is reduced as the ultradiscrete limit of the addition formulae for the level-three theta functions.
We also showed that the Hessian group, the group of linear automorphisms acting on the Hesse pencil, reduced to the dihedral group of degree three acting automorphically on the tropical Hesse pencil. 
By using the group structure on the tropical Hesse pencil, we can construct both integrable dynamical systems and solvable chaotic systems each of which is given by a piecewise linear map on the curve.

Although we only considered the linear automorphisms acting on the Hesse pencil in this paper, there exists the group of a wide class of automorphisms acting on the pencil. 
To investigate a tropical analogue of the Cremona group, the group of birational automorphisms acting on the Hesse pencil, is a further problem.

\section*{Acknowledgments}
This work was partially supported by Grants-in-Aid for Scientific Research, Japan Society for the Promotion of Science (JSPS), No. 22740100.

%######################################################%
%#################### BIBLIOGRAPHY ####################%
%######################################################%

\end{document}